# Local Edge Betweenness based Label Propagation for Community Detection in Complex Networks

Hamid Shahrivari Joghan
Computer Engineering and IT Department
Amirkabir University of Technology
Tehran, Iran
hamid.sh.j@aut.ac.ir

Alireza Bagheri
Computer Engineering and IT Department
Amirkabir University of Technology
Tehran, Iran
ar_bagheri@aut.ac.ir

*Nowadays, identification and detection community structures in complex networks is an important factor in extracting useful information from networks. Label propagation algorithm with near linear-time complexity is one of the most popular methods for detecting community structures, yet its uncertainty and randomness is a defective factor. Merging LPA with other community detection metrics would improve its accuracy and reduce instability of LPA. Considering this point, in this paper we tried to use edge betweenness centrality to improve LPA performance. On the other hand, calculating edge betweenness centrality is expensive, so as an alternative metric, we try to use local edge betweenness and present LPA-LEB (Label Propagation Algorithm Local Edge Betweenness). Experimental results on both real-world and benchmark networks show that LPA-LEB possesses higher accuracy and stability than LPA when detecting community structures in networks.*

Regular Research Papers

CSCI-ISNA

## I. INTRODUCTION

Modeling variant problems as complex networks helps clarifying other aspect of systems. Complex networks are wide-spread and can be found in many fields. Social networks can be represented as complex networks in which nodes indicate persons and edges show friendship relation between two persons. In biological networks which applies to biological systems complex networks can be used to model species units linked into a whole food web. The World Wide Web can be represented by web pages as nodes and hyperlinks as edges.

Detecting community segments represent meaningful intrinsic relations between entities of networks. Indeed, communities are subgraphs in which edges between sub-graph nodes are denser than edges between rest of the network, so the entities of community share more common property. For instance, communities detected in social networks represent real world friendship group or in WWW networks illustrate webpages in same field of information.

Although various algorithms we presented to detect community segments in complex network, none of them is comprehensive because detecting community structures is NP-complete. Scholars have tried various approaches for detecting communities such as heuristic methods, optimization problems or random algorithms, furthermore some algorithm are combination of several methods. We present LPA-LEB which is a random-based algorithm that uses local edge betweenness heuristic in order to improve accuracy.

LPA-LEB detects disjoint communities with more accuracy and stability. Same as LPA [1], LPA-LEB complexity in detecting communities of sparse graphs is near liner-time. Both traditional LPA and LPA-LEB are based on iterative calculation, but it is worthy to mention that Raghavan et al. claim that five iterations are sufficient to classify 95% of nodes correctly. Same as LPA, experiments show LPA-LEB could detect acceptable community structure in less than even 4 iterations. Moreover, experimental results indicate that LPA-LEB not only finds more accurate answers than LPA but also its results are more stable. The rest of this paper is organized as follows:

Related works are presented in section 2. Section 3 described some preliminaries including basic definitions, LPA presentation and describing local edge betweenness centrality. In Section 4, LPA-LEB is detailed and section 5 compares LPA-LEB against other algorithms on benchmark graphs and real world networks. At the end, conclusion and future works are mentioned.

## II. RELATED WORKS

Gregory presented LPAm which examined the label-propagation algorithm as an optimization problem [2]. However, it was prone to get stuck in poor local maxima in the modularity space. Therefore, a multistep greedy agglomerative algorithm (MSG) is proposed that can merge multiple pairs of communities at a time. After that, an advanced modularity-specialized label propagation (LPAm+) is proposed [3]. In another paper [4], Kang and Jia proposed an improved LPA algorithm called label propagation algorithm based on local similarity (LPALS) in which the similarity was used to be the weight value of the node labels.

Y. Xing et al. [5] proposed NIBLPA, which updates nodes label based on fixed node orders of label updated in the descending order of node importance. In another paper, R. Francisquini [6] tries to use a metaheuristic algorithm and introduces GA-LP, that its main idea is the local nature of the key operators of genetic algorithm. X. Zhang et al. [7] represent LPA-NI which is label propagation algorithm for community detection based on node importance and label influence. Zhao et al [8] proposed a novel algorithm for community detection called Label-Influence-Based (LIB) which selects a set of nodes as seeds and the label propagation procedure begins from the seeds. H. Lou et al [9] presented weighted coherent neighborhood propinquity (weighted-CNP) to calculate the probability that a pair of vertices is involved in a same community, in order to improve LPA accuracy.

Based on the mutual information of direct and indirect neighbors, N. Chen et al [10] introduced LPA-E which is capable of detecting both overlap and disjoint community. X. Zhang et al [11] presented LPALC which is also able to detect overlapping communities. LPALC is based on LPA and improves propagation process by choosing the nearest neighbor having a local cycle instead of choosing a neighbor randomly. J. Xie et al [12] proposed SLPA that unlike traditional LPA, holds more than one label for each node, so it is capable of detecting overlap community, too. Gregory put forward COPRA [13] which can detect overlapping community segments. Same as SLPA strategy, COPRA assigns multi labels to each node.

In another work, X. Zhang et al [14] introduced random-based algorithm (LPAc) based on edge clustering coefficient [15]. LPAc strategy is assumes that nodes label whose edge clustering coefficient is higher to be propagated preferentially. It is worthy to mention, LPAc is the most similar algorithm to LPA-LEB. There are different algorithms to detect communities, but this section just focuses on LPA based algorithm. For more details, many surveys compare variant approaches and algorithms for detecting both disjoint and overlap communities [16][17][18][19][20].

### III. PRELIMINARIES

A complex network is a graph represented as $G$ ($V$, $E$), in which $V$ is set of nodes and $E$ is set of edges connecting two nodes, also we assume that the graph is undirected and unweighted. We assume that $n$ represents size of $V$, $m$ represents size of $E$, $d$ represents average degree of nodes and $L_i$ represents label of node $i$. Graph represented as adjacency list named edges, also All indexes start from zero. Next, LPA and local edge betweenness are explained.

#### A. LPA

LPA is random-based algorithm and propagates labels of nodes at each iteration. Each node obtains a label with most frequency in its neighbors and after each iteration if all nodes label is most frequent label in its neighborhood then algorithm stops. In the worst accuracy, LPA holds whole graph as one community, so all nodes have same label. The LPA can be detailed as the following steps:

(1) Initialize unique labels to all nodes in the network and for a given node $x$, $L_x=x$.

(2) Rearrange the nodes in the network in a random order and set it to $X$.

(3) For each $x \in X$ chosen in that specific order, let $L_x$ obtain the label occurring with the highest frequency among neighbors.

(4) If every node has a label that the maximum number of their neighbors have, then stop the algorithm; go to (2).

(5) All nodes with same label are added to one community.

The complexity of LPA is $O(m)$ and in sparse graph is $O(n)$; however, LPA might divide a same graph in different communities in each execution, also its accuracy compared to other algorithms is low.

#### B. Local Edge Betweenness

Edge betweenness [21] is a measure for edge importance which indicates the number of shortest paths between pairs of vertices that paths through it. In case that, more than one shortest path excites between a pair of vertices, each path obtains equal weight, so the total weight of all of the paths is equal to one. Newman and Girvan [21] use edge betweenness in order to remove edges between communities but complexity of algorithm is $O(n.m^2)$; therefore, in large graph calculating edge betweenness is inefficient.

In another article [22], Steve Gregory introduces CONGO which use local edge betweenness in order to reduce complexity of Girvan-Newman algorithm. For calculating local edge betweenness of edge $e$, instead of counting all shortest path, just counts $h$ depth shortest path running along $e$. LPA-LEB uses 2-depth local edge betweenness to distinct between edges which are in same community and those are not. Complexity of calculating edge betweenness for all edges is $O(n.m)$, because it adapts BFS on each node to find all shortest paths between each nodes; however, $h$-depth local edge betweenness requires $h$-depth BFS, so 2-depth local edge betweenness complexity is $O((m/n)^2)$ and in sparse graph is $O(n)$ which is near liner-time.

### IV. THE PROPOSED ALGORITHM

LPA-LEB uses local edge betweenness in purpose to accurate and stabilize LPA. The main trait of LPE-LEB is limiting domain of probable assignable label for each node; indeed, it prevents nodes to obtain labels from neighbor nodes with high local edge betweenness. The following steps describe LPA-LEB in details:

(1) calculate 2-depth local edge betweenness of all edges.

(2) Initialize unique labels to all nodes in the network and for a given node $x$, $L_x=x$.

(3) Rearrange the nodes in the network in a random order and set it to $X$.

(4) For each $x \in X$ chosen in that specific order, let $L_x$ obtain the label occurring with the highest frequency among half+1 neighbor with littlest local edge betweenness.

(5) Rearrange the nodes in the network in a random order and set it to $X$.

(6) For each $x \in X$ chosen in that specific order, let $L_z$ obtains the label occurring with the highest frequency among neighbors.
(7) If every node has a label that the maximum number of their neighbors have, then stop the algorithm; go to (3).
(8) All nodes with same label are added to one community.

As mentioned previously, LPA-LEB uses 2-depth local edge betweenness. 1-depth local edge betweenness represents each node degree and it is inefficient to distinct between neighbor nodes label. 3-depth and deeper local edge betweenness give better accuracy; however, computing more depth causes more complexity. Thereby, LPA-LEB uses 2-depth local edge betweenness not only to keep complexity of algorithm near linear time in sparse graph, but also to obtain acceptable accuracy.

In step (4), each node searches half+1 of its neighbors with littlest local edge betweenness. If at most half+1 of the neighbors contain the same label $l$ then node belongs to community $l$. on the other hand, Newman and Girvan in their paper mention that the edge between two communities has big edge betweenness, so edges with little edge betweenness might be in the same community. Therefore, node $e$ tries to obtain label from half+1 of its neighbors with the littlest local edge betweenness, some of which would be in same community with probably $e$.

In step (6), each node obtains most frequent label from its whole neighbor because not necessarily edges with little local edge betweenness are in same community. Indeed, step (4) tries to create little accurate communities when neighbor labels are so variant in initial iterations and step (6) tries to accurate communities when communities become bigger in final iterations.

A remaining question is, if two labels both are most frequent labels, which of them would be assigned to node. In LPA one of them randomly would be chosen, but in LPA-LEB if this condition happens in step (6) then the label with littler local edge betweenness is assigned to node and in step (4), same as LPA, one of them randomly would be chosen.

Figure 1 shows local edge betweenness of Zachary's karate club [23] network. The edges inside community have smaller local edge betweenness than other edges which are between two communities. For instance, node 1 has 9 neighbors which neighbors possessing little local edge betweenness are in same community with node 1. Moreover, the edges between two communities would have less chance in algorithm because they have bigger value rather than its neighbors.

Figure 2 illustrates result of first iteration and clearly shows how a same label is spread through community, but still some nodes are not in their correct community. In Figure 3 four communities are obviously clarified; however, two nodes $a$ and

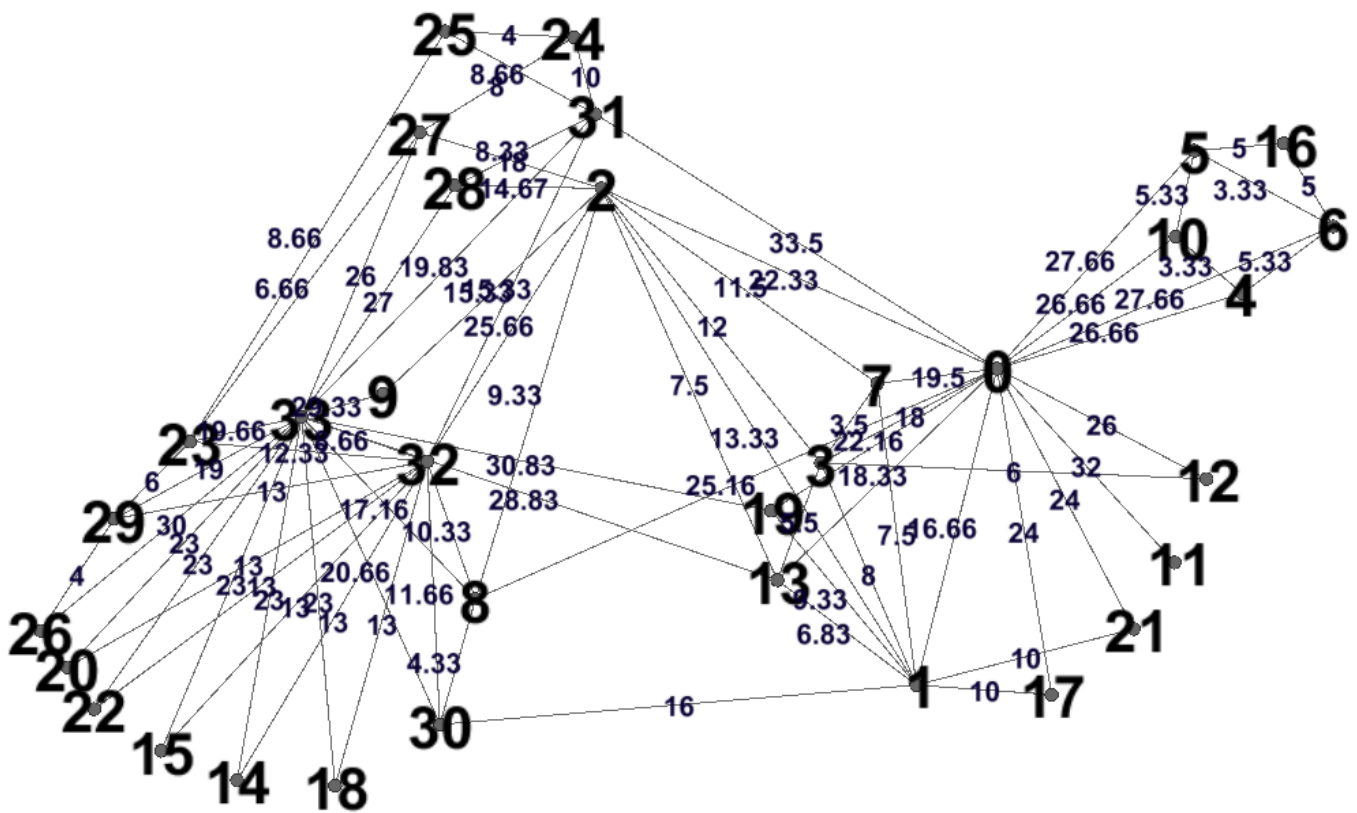

Fig. 1. Local edge betweenness of Zachary karate club which consists of 33 nodes and 78 edges. The edges between two communities possess bigger value than their neighbor.

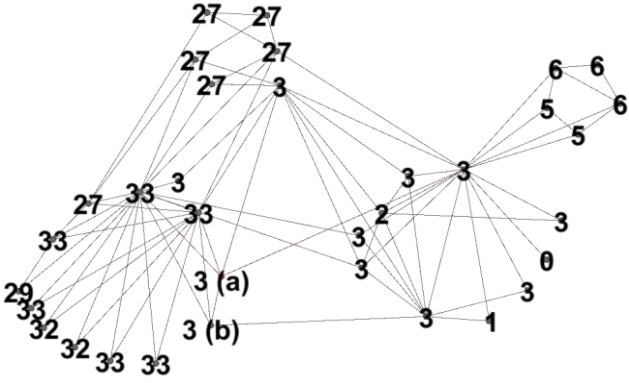

Fig. 2. Zachary karate club network after LPA-LEB first iteration. Communities are approximately clear but some nodes possess wrong label.

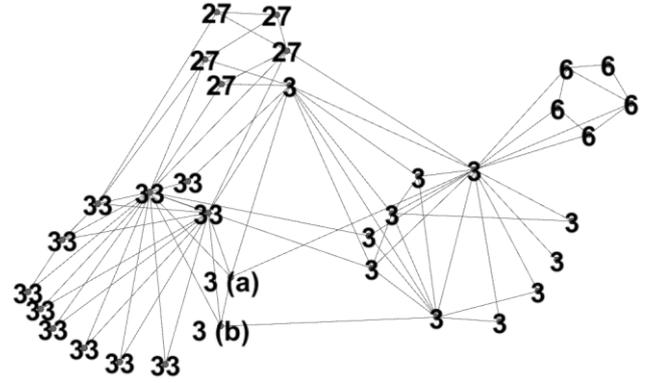

Fig. 3. Zachary karate club detected communities. These communities are result of LPA-LEB third iteration in which algorithm stopped.

*b* could have gained better label, yet communities approved stop criteria. This condition is evitable and its reason is randomness. In Figure 2, at the end of first iteration both these two nodes obtain label 3 and in next iteration node *b* saves its label because it is the most frequent label, but node *a* has two frequent label and accidently always choose label 3 instead of label 33.

Time complexity of steps (2), (3) and (5) is $O(n)$, time complexity of step (6) is $O(d.n)$, time complexity of both steps (1) and (3) are $O(d^2.n)$, so time complexity of whole algorithm is $O(d^2.n)$ and for sparse graph is $O(n)$ which is near linear time. The algorithm requires $O(m)$ space for saving network as adjacency list and local edge betweenness of each edge; besides, LPA-LEB needs $O(d)$ temporary space for calculating 2-depth local edge betweenness.

## V. EXPERIMENTS

In order to verify the performance of LPA-LEB, we use synthesized networks and real-world networks to measure accuracy and stability of LPA-LEB. For comparison, LPA and LPAc are chosen, because LPA-LEB main purpose is improving LPA accuracy and stability and LPAc is most similar algorithm to LPA-LEB. We use two different stop criteria for our algorithm: LPA-LEB which use traditional LPA stop criteria and LPA-LEB with maximum 4 iteration. All the algorithms are implemented in C++ and simulations are carried out in a desktop PC with Inter Core CPU i5-4460 @ 3.20 GHz and 4 GB memory under Windows 10.

As mentioned above, experiments are both on real-world and synthesized networks. Table 1 represents real-world networks and their properties. The modularity [24] value is used to appraise quality of detected communities in real-world networks. Synthetic network used in experiments is generated by LFR [25] benchmark which generates random networks with ground truth communities; thereby, in order to evaluate algorithm accuracy NMI [26] measurement is used, for LFR presents ground truth communities.

The algorithms are executed 1000 times and maximum iteration number is set to 50. Average result, best result, worst result, the variance of results and the average number of detected communities are represented in Table 2. Experiments indicate LPA-LEB detect communities almost more accurate than LPA and LPAc. In Karate network both LPA and LPAc might detect whole network as community, but this condition never happened with LPA-LEB. Besides, variance of results in LPA-LEB is lesser than LPA and LPAc that imply LPA-LEB is more stable.

As mentioned before, LPA-LEB could detect accurate community structure even with less than 5 iterations. Table 3 represents LPA-LEB performance when its maximum iteration number is 4. Almost in all networks LPA-LEB acts acceptable, but in Power networks its accuracy decreases dramatically. Indeed, if sizes of communities are so big or order of $O(n)$, it is not recommended use limited iteration number especially with little value, because it prevents algorithm to enlarge communities size and resulted communities would be smaller than their actual size.

Figure 4 illustrates comparative experiments on artificial networks, which are generated by LFR. The algorithms are executed 100 times and maximum iteration number is set to 50 and average NMI is calculated. Until mutual parameters is lesser than 0.5, LPA-LEB detects communities a bit accurate than LPA and LPAc; however, when mutual parameter exceeds 0.5, performance of LPA-LEB reduces dramatically, for community structures are not clearly separated. It is worthy to mention when maximum iteration is set 4 then LPA-LEB results are more accurate for network with mutual parameter set 0.6, because as pointed before, in this network communities are not lucid separated and LPA-LEB is prevented to merge more communities together.

Table 1. the number of nodes, edges and average degree of real-world networks

| Network | Description | Nodes | Edges | Average Degree |
|---|---|---|---|---|
| Karate | Zachary's karate club [23] | 34 | 78 | 4.58 |
| Dolphin | Dolphin social network [27] | 62 | 159 | 5.12 |
| Football | American College football [21] | 115 | 613 | 10.66 |
| Power | Power grid [28] | 4941 | 6594 | 2.66 |

Table 2. Comparative experiment of LPA-LEB with LPA and LPAc on real-world networks. Average result, average detected communities, best result, worst result and variance of results are represented. * means that algorithm assumed whole network as community structure

| Network | Average modularity | | | Average communities count | | | variance | | | Best modularity | | | Worst modularity | | |
|---|---|---|---|---|---|---|---|---|---|---|---|---|---|---|---|
| | LPA-LEB | LPA | LPAc | LPA-LEB | LPA | LPAc | LPA-LEB | LPA | LPAc | LPA-LEB | LPA | LPAc | LPA-LEB | LPA | LPAc |
| Karate | **0.3906** | 0.3455 | 0.2939 | **3.15** | 2.53 | 2.94 | **0.0020** | 0.0084 | 0.0042 | **0.4155** | **0.4155** | 0.3717 | **0.1120** | * | * |
| Dolphin | **0.5152** | 0.4819 | 0.4897 | 5.11 | 3.96 | **5.51** | **0.0001** | 0.0025 | 0.0006 | 0.5264 | **0.5267** | 0.5246 | **0.4350** | 0.3017 | 0.3322 |
| Football | **0.5980** | 0.5899 | 0.5791 | **11.73** | 10.7 | 11.58 | **0.0000** | 0.0001 | 0.0003 | **0.6045** | **0.6045** | 0.6033 | **0.5330** | 0.5299 | 0.5148 |
| Power | 0.7844 | **0.8002** | 0.6860 | 528.6 | 501.97 | **995.01** | 0.0000 | 0.0000 | 0.0000 | 0.7794 | **0.8126** | 0.6933 | **0.7887** | 0.7885 | 0.6797 |

Table 3. comparative experiments based on modularity of LPA-LEB with maximum 4 iteration with LPA-LEB on real-world networks

| Network | Average modularity | | Average communities count | | variance | | Best modularity | | Worst modularity | |
|---|---|---|---|---|---|---|---|---|---|---|
| | LPA-LEB 4 iteration | LPA-LEB | LPA-LEB 4 iteration | LPA-LEB | LPA-LEB 4 iteration | LPA-LEB | LPA-LEB 4 iteration | LPA-LEB | LPA-LEB 4 iteration | LPA-LEB |
| Karate | 0.3882 | **0.3906** | **3.26** | 3.15 | **0.0019** | 0.0020 | **0.4174** | 0.4155 | 0.1120 | 0.1120 |
| Dolphin | 0.5118 | **0.5152** | **5.62** | 5.11 | 0.0001 | 0.0001 | **0.5277** | 0.5264 | 0.4350 | 0.4350 |
| Football | 0.5959 | **0.5980** | 11.8 | 11.73 | 0.0001 | **0.0000** | 0.6045 | 0.6045 | 0.5568 | 0.5330 |
| Power | 0.7115 | **0.7844** | **835.2** | 528.66 | 0.000 | 0.0000 | 0.7183 | 0.7794 | **0.7054** | 0.7887 |

Table 4. different modularity resulted with deeper local edge betweenness. Last column named n-depth indicates effect of using edge betweenness instead of local edge betweenness.

| Network | 2-depth | | 3-depth | | n-depth | |
|---|---|---|---|---|---|---|
| | LPA-LEB | LPA-LEB 4 iteration | LPA-LEB | LPA-LEB 4 iteration | LPA-LEB | LPA-LEB 4 iteration |
| Karate | 0.3906 | 0.3882 | 0.3991 | 0.3982 | **0.4012** | 0.4003 |
| Dolphin | **0.5152** | 0.5118 | 0.5141 | 0.5116 | 0.5094 | 0.5092 |
| Football | 0.5980 | 0.5959 | **0.6016** | 0.5992 | 0.6002 | 0.5985 |
| Power | 0.7844 | 0.7115 | 0.7907 | 0.7180 | **0.8062** | 0.7295 |

LPA-LEB uses 2-depth local edge betweenness to keep time complexity near linear time, but it is possible to use deeper local edge betweenness even based on all shortest path. comparison experiments on different local edge betweenness is presented in Table 4. Although deeper local edge betweenness might detect community structure more accurately, yet 2-depth local edge betweenness with near linear complexity represents satisfied answers even in some cases better.

## VI. CONCLUSION

This article introduces random-based LPA-LEB which is nonparametric algorithm for detecting community structure in complex networks. The time and space complexity of this algorithm is near liner in sparse networks. LPA-LEB has tangible improvement both considering accuracy and stability compared to LPA and LPAc. Various experiments represented in this paper confirm that accuracy of LPA-LEB is acceptable and it has good performance in detecting community structures in both real-world networks and artificial networks. Experiments also show LPA-LEB could detect communities accurately even in large scale graph with 100,000 nodes.

## VII. FUTURE WORKS

In this paper, all experiments are on undirected and unweighted networks, but it does not consider that LPA-LEB is not capable of detecting communities in directed or weighted graphs. The local edge betweenness used as improver metric is computable in directed or weighted graph, so it seems possible to generalize LPA-LEB to detect directed or weighted communities for future works. Moreover, future works might include scaling LPA-LEB up in distributed system which is popular nowadays.

## ACKNOWLEDGMENT

We appreciate Saeed Shahrivari for reviewing this article and thank Navid Sedighpour for explaining LPA also gratuity Rasoul Mousavi because of his narrow ideas which improved quality of this article.

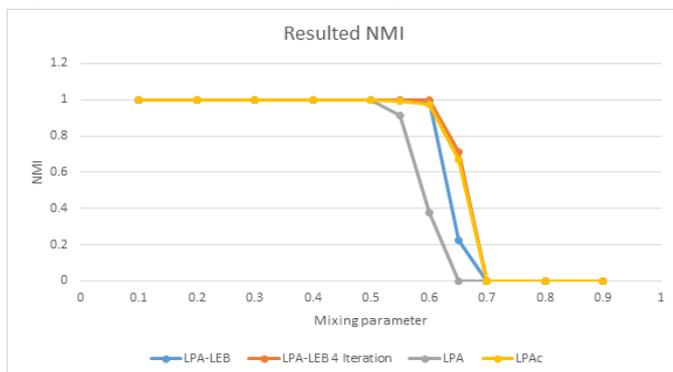

Fig. 4. LFR network with N=1000, k=20, maxk=50, minc=10, maxc=50, om=0, t1=2 and t2=1. Blue curve represents LPA-LEB, red curve represents LPA-LEB with maximum 4 iteration, LPA is gray curve and yellow curve is LPAc.